\documentclass
[aps,preprint,showkeys,a4paper,tightenlines,superscriptaddress]{revtex4}%
\usepackage{eurosym}
\usepackage{amsfonts}
\usepackage{amsmath}
\usepackage{amssymb}
\usepackage{graphicx}
\usepackage{subfig}
\usepackage{caption}%
\providecommand{\U}[1]{\protect\rule{.1in}{.1in}}
\providecommand{\U}[1]{\protect\rule{.1in}{.1in}}
\newtheorem{theorem}{Theorem}
\newtheorem{acknowledgement}[theorem]{Acknowledgement}

\begin{document}
\title{The role of interfacial friction on the peeling of thin viscoelastic tapes}
\author{M. Ceglie}
\affiliation{Department of Mechanics, Mathematics and Management, Politecnico of Bari, V.le
Japigia, 182, 70126, Bari, Italy}
\author{N. Menga}
\email{nicola.menga@poliba.it}
\affiliation{Department of Mechanics, Mathematics and Management, Politecnico of Bari, V.le
Japigia, 182, 70126, Bari, Italy}
\affiliation{Imperial College London, Department of Mechanical Engineering, Exhibition
Road, London SW7 2AZ}
\author{G. Carbone}
\affiliation{Department of Mechanics, Mathematics and Management, Politecnico of Bari, V.le
Japigia, 182, 70126, Bari, Italy}
\affiliation{Imperial College London, Department of Mechanical Engineering, Exhibition
Road, London SW7 2AZ}
\begin{abstract}
We study the peeling process of a thin viscoelastic tape from a rigid
substrate. Two different boundary conditions are considered at the interface
between the tape and the substrate: stuck adhesion, and relative sliding in
the presence of frictional shear stress. In the case of perfectly sticking
interfaces, we found that the viscoelastic peeling behavior resembles the
classical Kendall behavior of elastic tapes, with the elastic modulus given by
the tape high-frequency viscoelastic modulus. Including the effect of
frictional sliding, which occurs at the interface adjacent to the peeling
front, makes the peeling behavior strongly dependent on the peeling velocity.
Also, at sufficiently small peeling angles, we predict a tougher peeling
behavior than the classical stuck cases. This phenomenon is in agreement with
recent experimental evidences indicating that several biological systems (e.g.
geckos, spiders) exploit low-angle peeling to control attachment force and locomotion.

\end{abstract}
\maketitle

\section{Introduction}

The ability to control the mechanical behavior of real interfaces is one of
the most challenging topic in modern industrial engineering, as witnessed by
the effort made in the last decades in elastic
\cite{Hyun2004,Campana2008,YangPersson2008,Pastewka2016,menga2016,Muser2017,menga2018bis,menga2019geom}%
, viscoelastic
\cite{Persson2001,Persson2004,Scaraggi2015,Menga2014,menga2016visco,Menga2018,Zhang2020,Menga2021}
and adhesive
\cite{Carbone2008,Martina2012,Pastewka2014,DiniMedina,Rey2017,MCD2018,MCD2018corr,Menga2019}
contact mechanics studies. In this respect, the functionality of several
real-life devices such as, for instance, pressure sensitive adhesives, modern
touch screens, biomedical wound dressing and Band-Aids, is tightly connected
to the possibility of tailoring the interfacial adhesion between these systems
and the mating surfaces. Therefore, the attachment/detachment performances of
such interfaces are of primary concern. Among the possible detachment
mechanisms, peeling is usually regarded as one of the most important
\cite{Zhu2021} as it offers the chance to independently variate both the
peeling load $P$ and the peeling angle $\theta$. Due to its reliability, also
the adhesive performance of industrial interfaces is often controlled by means
of opportunely designed peeling tests \cite{Creton2016}. For these reasons, a
deep understanding of the physics behind peeling processes is of primary
importance in modern engineering.

The basic understanding of the peeling behavior of thin elastic tapes is
nowadays well established. Indeed, Kendall's \cite{Kendall1975} studies set
the theoretical framework to investigate the peeling evolution by relying on
energy balance of the system. His pioneering results were valid under specific
hypothesis, such as the linear elasticity of the tape, perfect interfacial
backing between the tape and the substrate. However, in the last decades, most
of the original assumptions have been relaxed in successive studies. This is
the case, for instance, of the effect of bending stiffness in thick tapes,
which has been investigated in Refs \cite{Kendall1973a,Kendall1973b} in the
presence of viscous losses occurring in the adhesive layer and in the tape. In
Refs \cite{Afferrante2016,Pierro2020}, the effect of the substrate
viscoelasticity on the peeling behavior has been studied to show that the
steady detachment speed can be tuned under specific conditions and ultra-tough
peeling may occur at low peeling angles. The effect of the tape
viscoelasticity has been investigated in Ref.
\cite{Derail1997,Zhou2011,Chen2013,Peng2014}, where it has been empirically
shown that, in a limited range of peeling velocities \cite{Zhu2021}, the
peeling force $P\approx kV^{n}$.

Also the tape-substrate interfacial interactions play a fundamental role in
determining the evolution of the peeling process. Kendall showed that
interfacial sliding between the tape and the substrate may trigger cyclic
detachment and re-attachment of the tape \cite{Kendall1975b,Kendall1978}.
Nonetheless, such a relative sliding represents an additional source of energy
dissipation due to frictional shear stress occurring at the interface
\cite{Newby1995,Zhang Newby1997}. This has been investigated in details in the
case of elastic tapes peeled away from rigid substrates under frictionless
\cite{Wang2007} and frictional \cite{Begley2013} sliding. Theoretical and
experimental investigations have concluded that, as expected, the presence of
frictional sliding at the interface leads to significantly higher peeling
force, which may theoretically diverge at vanishing peeling angle. It has been
suggested that the enhancement of peeling force caused by the energy
dissipated in frictional sliding plays a key role in biological applications.
Indeed, several studies have identified this mechanism as one of the possible
sources of the superior adhesive performance and the ability to switch between
firm attachment and effortless detachment of geckos
\cite{Autumn2006a,Autumn2006b,Tian2006}, insects
\cite{Labonte2015,Endlein2013b} and frogs \cite{Endlein2013a}, together with
the hierarchical configuration of the pad/toes fibrils
\cite{Gao2005,Lee2012,Zhao2013}, and the V-shaped peeling scheme often
occurring at the interface
\cite{Heepe2017,Afferrante2013,mengaVpeeling,mengaVpeelingthin}. Specifically,
in the presence of interfacial sliding, the enhanced peeling force arises due
to the combination of three main physical mechanisms \cite{Labonte2016}: (i)
tape pre-strain due to partial sliding occurring during detachment
\cite{Chen2009,Williams1993}; (ii) the friction losses associated with the
slip at the interface \cite{Amouroux2001}; (iii) the viscous dissipation
\cite{Labonte2015}. Interestingly, in wet contact case, the last contribution
is usually ascribed to the viscous shear in the interfacial thin fluid film
formed by the pad secretion \cite{Federle2002,Federle2004}; however, a
comprehensive model on the effect of the pad tissue bulk viscoelasticity
\cite{Gorb2000,Puthoff2013} on the overall frictional peeling is lacking.

In this study, we present a theoretical model of the behavior of a thin
viscoelastic tape peeled away from a rigid substrate. Specifically, we aim at
investigating the combined effect of frictional interfacial sliding occurring
during the detachment process and the energy dissipation associated with the
viscoelastic behavior of the tape. In order to shed light on the interplay
between these mechanisms of energy dissipation, we consider two different
configurations: (i) firstly, we focus on stuck conditions, where a rigid
constraint avoids any possible interfacial displacement between the tape and
the substrate, so that no additional energy contribution is present at
interface beside the change in the energy of adhesion; then (ii)\ we consider
the sliding case, where energy dissipation occurs due to frictional\ relative
sliding in the tape elongated region. Our findings may be of help to estimate
the effect of the bulk viscoelasticity on the overall peeling behavior of
bioinspired and natural systems (e.g. biological fibrils, industrial polymeric
tapes, etc.) in the presence of interfacial frictional sliding.

\section{Formulation}

In this section, we present the mathematical model for the viscoelastic
peeling assuming two different conditions at the interface between the tape
and the rigid substrate: (i) stuck adhesion, where tape normal and tangential
displacements are inhibited; (ii) relative frictional sliding, i.e. tape
normal displacements are inhibited but tangential sliding occurs, which is
opposed by frictional shear stresses. In both cases, the problem formulation
builds on energy balance. We focus on a linear viscoelastic material with a
single relaxation time $t_{c}$. Further, we neglect any dynamic effect during
the tape detaching process (\textit{i.e.} we assume that the peeling front
velocity is steady and far lower than the speed of sound in the tape).

\subsection*{Stuck interface}

\begin{figure}[ptbh]
\begin{center}
\includegraphics[width=0.5\textwidth]{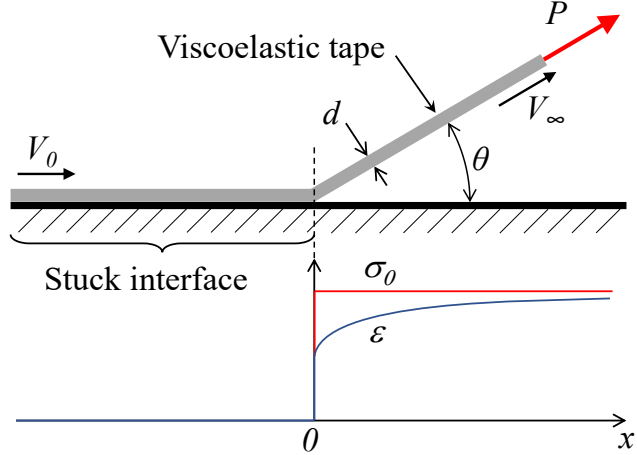}
\end{center}
\caption{The scheme of the peeling process of a thin viscoelastic layer from a
rigid substrate in the presence of stuck adhesion at the interface, so that no
relative sliding occurs. In the lower part, qualitative diagrams of the tape
stress $\sigma$ (red) and deformation $\varepsilon$ (blue) are shown.}%
\label{fig1}%
\end{figure}

Consider a viscoelastic tape of thickness $d$ and transverse width $b$, baked
onto a rigid substrate with no relative sliding at the interface. As shown in
Fig. \ref{fig1}, the tape is peeled away at an angle $\theta$ under a constant
force $P$. We assume the peeling front moves on the left at a constant
velocity $V_{0}$ relative to the substrate. We observe the peeling process in
the peeling front, so that the substrate moves on the right at speed $V_{0}$
as shown in Fig. \ref{fig1}.

Under steady state conditions, the energy balance per unit time of the
viscoelastic tape is given by
\begin{equation}
W_{E}+W_{I}+W_{S}=0 \label{eq:equilibrium 1}%
\end{equation}
where $W_{E}$ is the work per unit time of the external forces acting on the
tape, $W_{I}$ is work per unit time done by tape internal stresses, which
takes into account for both the change in the stored elastic and viscous
energy dissipation in the tape, and $W_{S}$ is the work per unit time done by
interfacial forces. Notably, in this formulation we neglect any other source
of energy dissipation, such as acoustic or thermal emissions.

The term $W_{E}$ in Eq. (\ref{eq:equilibrium 1}) can be calculated considering
the external forces acting on the system, which are the remote load $P$ acting
on the detached tape tip, and the corresponding opposite substrate reaction
force $-P\cos\theta$. We have
\begin{equation}
W_{E}=PV_{\infty}-PV_{0}\cos\theta=\sigma_{0}V_{0}bd\left(  1+\frac{\sigma
_{0}}{E_{0}}-\cos\theta\right)  \label{eq:external power}%
\end{equation}
where we defined $\sigma_{0}=P/bd$, and $E_{0}$ is the low frequency
viscoelastic modulus. Moreover, the mass balance of the tape gives $V_{\infty
}=V_{0}(1+\sigma_{0}/E_{0})$. Notably, in Eq. (\ref{eq:external power}) we
assumed that the tape tip (where the force $P$ is applied) is located
sufficiently far from the peeling front, so that complete viscoelastic
relaxation occurs in the detached tape.

The surface term $W_{S}$ in Eq. (\ref{eq:equilibrium 1}) represents the energy
per unit time associated with the rupture of interfacial adhesive bond, with
$\Delta\gamma$ being the work of adhesion needed to detach a unit surface of
the tape from the substrate. Although $\Delta\gamma$ may also depend on the
viscoelastic energy dissipation occurring very close to the crack tip (i.e.
small-scale viscoelasticity), for the sake of simplicity, here we neglect such
effects, thus assuming a constant $\Delta\gamma$ value. Nonetheless, this
effect could be straightforwardly introduced in the present model by
opportunely modifying the value of $\Delta\gamma$ as described in Refs.
\cite{Carbone2005a,Carbone2005b}. Therefore, we can write%
\begin{equation}
W_{S}=-\Delta\gamma bV_{0} \label{eq:adhesive power}%
\end{equation}

As mentioned above, the term $W_{I}$ in Eq. (\ref{eq:equilibrium 1}) takes
into account for both the elastic energy stored in the tape, and the bulk
energy dissipation occurring due to viscoelastic creep in the detached strip.
Moreover, observing that the bending stiffness of the tape depends on the
third power of thickness $d$, and considering that we focus on very thin
tapes, the bending contribution to$\ W_{I}$ can be neglected (see also Refs.
\cite{mengaVpeeling,mengaVpeelingthin}). Hence we write%
\begin{equation}
W_{I}=-V_{0}bd\int_{-\infty}^{+\infty}\sigma(x)\varepsilon^{\prime}(x)\,dx
\label{Wint_1}%
\end{equation}
where $\varepsilon^{\prime}(x)$ is the spatial derivative of the strain
$\varepsilon\left(  x\right)  $. Note that, in Eq. (\ref{Wint_1}), we used
$\dot{\varepsilon}(x)=V_{0}\varepsilon^{\prime}(x)$, with $\dot{\varepsilon
}(x)$ being the time derivative of $\varepsilon\left(  x\right)  $.

Since in this section we assume no interfacial sliding between the adhering
tape and the rigid substrate, the stress distribution in the viscoelastic tape
is given by $\sigma(x)=\sigma_{0}H(x)$, with $H(x)$ being the Heaviside step
function (see the diagram in Fig. \ref{fig1}). Moving from the linear
viscoelastic constitutive equation \cite{christensen}, in the framework of
steady state conditions, the deformation field can be calculated as%
\begin{equation}
\varepsilon(x)=\int_{-\infty}^{x}J(x-s)\sigma^{\prime}(s)\,ds
\label{eq:constitutive}%
\end{equation}
where\ $J(x)$ is the spatial transformation of the viscoelastic creep function
given by%
\begin{equation}
J(x)=H\left(  x\right)  \left[  \frac{1}{E_{0}}-\frac{1}{E_{1}}\exp\left(
-\frac{x}{\lambda}\right)  \right]  \label{creep}%
\end{equation}
where $\lambda=V_{0}t_{c}$ is the relaxation length, and $E_{1}^{-1}%
=E_{0}^{-1}-E_{\infty}^{-1}$ with $E_{0}$ and $E_{\infty}$ being the low and
very high frequency viscoelastic moduli, respectively.

For the case at hand, Eq. (\ref{eq:constitutive}) gives $\varepsilon
(x)=\sigma_{0}J(x)$, which substituting into Eq. (\ref{Wint_1}), after some
algebra, gives
\begin{equation}
W_{I}=-V_{0}bd\sigma_{0}^{2}\left(  \frac{1}{E_{0}}-\frac{1}{2E_{\infty}%
}\right)  =-\frac{V_{0}bd\sigma_{0}^{2}}{2}\left(  \frac{1}{E_{0}}+\frac
{1}{E_{1}}\right)  \label{Wint_bis}%
\end{equation}
where we used $\int_{-\infty}^{+\infty}\delta(x)H\left(  x\right)  =1/2$,
which gives $\varepsilon(0)=\frac{1}{2}\left[  \varepsilon\left(
0^{-}\right)  +\varepsilon\left(  0^{+}\right)  \right]  =\frac{1}%
{2}\varepsilon\left(  0^{+}\right)  $. Note that the viscoelastic dissipated
energy per unit time is
\begin{equation}
D_{\mathrm{s}}=\frac{V_{0}bd\sigma_{0}^{2}}{2E_{1}}
\label{viscoelastic energy dissipation 1}%
\end{equation}
Finally, substituting Eqs. (\ref{eq:external power},\ref{eq:adhesive power}%
,\ref{Wint_bis}) into Eq. (\ref{eq:equilibrium 1}) we have%
\begin{equation}
\frac{\sigma_{0}^{2}}{2E_{\infty}}+\sigma_{0}(1-\cos\theta)=\frac{\Delta
\gamma}{d} \label{final}%
\end{equation}
which represents the peeling equilibrium condition. Interestingly regardless
of the peeling velocity $V_{0}$, Eq. (\ref{final}) is identical to the Kendall
equation with the elastic modulus given by the high frequency viscoelastic
modulus $E_{\infty}$. Notice that Eq. (\ref{final}) can be rephrased as%
\begin{equation}
\frac{1}{2}\left(  \frac{\sigma_{0}}{E_{0}}\right)  ^{2}+\kappa\frac
{\sigma_{0}}{E_{0}}(1-\cos\theta)=\kappa\frac{\Delta\gamma}{E_{0}d}%
\end{equation}
where we have defined $\kappa=E_{\infty}/E_{0}$. For $\theta=0$, we get%
\begin{equation}
\sigma_{K}=\sqrt{\frac{2\kappa E_{0}\Delta\gamma}{d}}=\sqrt{\frac{2E_{\infty
}\Delta\gamma}{d}} \label{sigma_1}%
\end{equation}
so that, in this case, the peeling is much more tough than in the (low
frequency) elastic case as the effective work of adhesion is $\kappa$-times
larger than $\Delta\gamma$.

In order to explain the appearance of the high-frequency viscoelastic modulus
$E_{\infty}$ in Eq. (\ref{final}) we note that, because of the stuck condition
assumption (no relative sliding at the tape-substrate interface), the tape is
subjected to an abrupt stretching in the peeling section (see Fig.
\ref{fig1}). For this reason, regardless of the peeling velocity $V_{0}$, the
material response close to the peeling front is governed by the high-frequency
viscoelastic response, which makes the tape locally behave as a perfectly
elastic material with elastic modulus $E_{\infty}$.

Notably, in real conditions, the abrupt change of the tape stress during
peeling would be smoothed, as it must occur on a finite length scale across
the peeling section. Since the size of this "process zone" can be estimated of
order unity of the tape thickness $d$ (see also Refs
\cite{Lin2007,mengaVpeeling}), the tape excitation frequency during peeling is
$\omega\approx V_{0}/d$, so that at very low peeling velocities, i.e. when
$V_{0}\ll d/t_{c}$, the tape response would be governed by the low-frequency
viscoelastic modulus $E_{0}$. However, since we usually expect that $V_{0}\gg
d/t_{c}$, this would not qualitatively affect the physical picture of the
peeling behavior provided so far.

\subsection*{Frictional sliding interface}

\begin{figure}[ptbh]
\begin{center}
\includegraphics[width=0.5\textwidth]{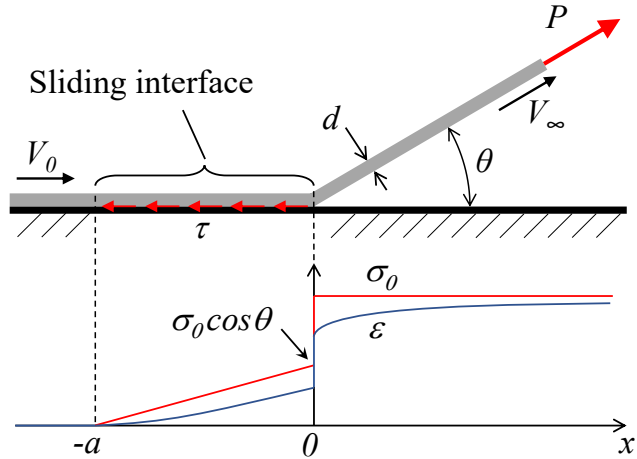}
\end{center}
\caption{The scheme of the peeling process of a thin viscoelastic layer from a
rigid substrate in the presence of relative sliding at the interface. Notably,
$\tau$ is the frictional shear stress. In the lower part, qualitative diagrams
of the tape stress $\sigma$ (red) and deformation $\varepsilon$ (blue)\ are
shown.}%
\label{fig2}%
\end{figure}

The discussion provided in the previous section is based on the assumption
that the tape firmly sticks to the rigid substrate, and the tangential
component of the peeling force $P$, remotely acting on the tape tip, is
locally balanced by a point reaction force acting in the peeling section.
However, it has been shown that a certain amount of relative sliding occurs in
real interfaces \cite{Newby1995,Amouroux2001,Collino2014}, so that the
tangential component of the peeling force $P$ is balanced by the frictional
shear stresses arising at the interface between the tape and the rigid
substrate. In this case, assuming a uniform interfacial shear stress $\tau$, a
portion of the adhering tape of length $a=P\cos\theta/\tau b$ is gradually
stretched and slides against the substrate. Such a physical scenario is shown
in Fig. \ref{fig2}.

The work per unit time\ done by interfacial frictional stresses is%
\begin{equation}
W_{T}=-\int_{-a}^{0}V_{S}(x)\tau bdx=-\tau bV_{0}\int_{-a}^{0}\varepsilon(x)dx
\label{Wf}%
\end{equation}
where $V_{S}\left(  x\right)  =V_{0}\varepsilon\left(  x\right)  $ is the
sliding velocity distribution at the interface. Of course, both the stress and
deformation distributions along the tape are modified due to the presence of
the tangential tractions $\tau$. Indeed, we have%
\begin{align}
\sigma\left(  x\right)   &  =\frac{\tau}{d}\left(  x+a\right)  ;\qquad-a\leq
x<0\nonumber\\
\sigma\left(  x\right)   &  =\sigma_{0};\qquad x>0 \label{sigma}%
\end{align}
where $\sigma_{0}cos\theta=\tau a/d$. Similarly,\ from Eq. (\ref{sigma}),
recalling, Eq. (\ref{eq:constitutive}), one obtains%
\begin{align}
\varepsilon(x)  &  =\frac{\tau}{E_{0}d}(a+x)-\frac{\tau\lambda}{E_{1}d}\left[
1-\exp\left(  -\frac{x+a}{\lambda}\right)  \right]  ;\qquad-a\leq
x<0\nonumber\\
\varepsilon(x)  &  =\frac{\sigma_{0}}{E_{0}}-\frac{\sigma_{0}}{E_{1}}\left\{
1-\frac{\tau a}{\sigma_{0}d}+\frac{\tau\lambda}{\sigma_{0}d}\left[
1-\exp\left(  -\frac{a}{\lambda}\right)  \right]  \right\}  \exp\left(
-x/\lambda\right)  ;\qquad x>0 \label{eps}%
\end{align}

Recalling Eq. (\ref{Wf}) and using Eqs. (\ref{sigma}, \ref{eps}) we have%
\begin{equation}
W_{T}=-bdV_{0}\frac{\sigma_{0}^{2}}{2E_{0}}\cos^{2}\theta\left\{
1-2\frac{\kappa-1}{\kappa}\frac{\lambda}{a}\left[  1-\frac{\lambda}{a}%
+\frac{\lambda}{a}\exp\left(  -\frac{a}{\lambda}\right)  \right]  \right\}
\label{friction}%
\end{equation}
In Eq. (\ref{friction}), we note that for $a/\lambda\rightarrow\infty$ we get
$W_{T}\rightarrow-\frac{1}{2}bdV_{0}(\sigma_{0}^{2}/E_{0})\cos^{2}\theta$,
which involves the low frequency modulus $E_{0}$; whereas, for $a/\lambda
\rightarrow0$ we get $W_{T}\rightarrow-\frac{1}{2}bdV_{0}(\sigma_{0}%
^{2}/E_{\infty})\cos^{2}\theta$, which involves the high frequency modulus
$E_{\infty}$. Moreover, $W_{T}\left(  a/\lambda\rightarrow\infty\right)
=\kappa$ $W_{T}\left(  a/\lambda\rightarrow0\right)  $.

This time, the work per unit time done by tape internal stresses is%
\begin{align}
W_{I}  &  =-bdV_{0}\int_{-\infty}^{+\infty}\sigma\left(  x\right)
\varepsilon^{\prime}\left(  x\right)  dx=\nonumber\\
&  -bdV_{0}\frac{\sigma_{0}^{2}}{2E_{0}}\left(  \cos^{2}\theta\left\{
1-2\frac{\kappa-1}{\kappa}\frac{\lambda}{a}\left[  \frac{\lambda}{a}-\left(
1+\frac{\lambda}{a}\right)  \exp\left(  -\frac{a}{\lambda}\right)  \right]
\right\}  \right. \nonumber\\
&  \left.  -\frac{1}{\kappa}\left(  1-\cos^{2}\theta\right)  -2\frac{\kappa
-1}{\kappa}\left\{  1-\cos\theta\left[  1-\frac{\lambda}{a}+\frac{\lambda}%
{a}\exp\left(  -\frac{a}{\lambda}\right)  \right]  \right\}  \right)
\label{WintSliding}%
\end{align}

Finally, recalling that, in this case, Eq. (\ref{eq:equilibrium 1}) modifies
in%
\begin{equation}
W_{E}+W_{I}+W_{S}+W_{T}=0 \label{equil_2}%
\end{equation}
and using Eqs. (\ref{eq:external power},\ref{eq:adhesive power},\ref{friction}%
,\ref{WintSliding}) into Eq. (\ref{equil_2}), the final peeling equilibrium
equation for a viscoelastic tape in the presence of frictional sliding at the
interface is given by%

\begin{gather}
\frac{\sigma_{0}^{2}}{2E_{0}}\left\{  \left(  1-\cos^{2}\theta\right)
-\frac{\kappa-1}{\kappa}\left(  1-\cos\theta\right)  \left(  1+2\cos
\theta\left[  \frac{\lambda}{a}\left(  1-\exp\left(  -\frac{a}{\lambda
}\right)  \right)  -\frac{1}{2}\right]  \right)  \right\} \nonumber\\
+\sigma_{0}(1-\cos\theta)-\frac{\Delta\gamma}{d}=0 \label{final_2}%
\end{gather}

\section{Numerical results}

Let us introduce the following dimensionless parameters:\ $\tilde{\sigma}%
_{0}=\sigma_{0}/E_{0}$, $\tilde{\tau}=\tau/E_{0}$, $\Delta\tilde{\gamma
}=\Delta\gamma/\left(  E_{0}d\right)  $ and $\tilde{V}_{0}=V_{0}t_{c}/d$. Note
that $a/\lambda=\tilde{\sigma}_{0}\cos\theta/\left(  \tilde{V}_{0}\tilde{\tau
}\right)  $. Therefore, Eq. (\ref{final_2}) becomes%
\begin{gather}
\frac{1}{2}\tilde{\sigma}_{0}^{2}\left\{  \left(  1-\cos^{2}\theta\right)
-\frac{\kappa-1}{\kappa}\left(  1-\cos\theta\right)  \left(  1+2\cos
\theta\left[  \frac{\tilde{V}_{0}\tilde{\tau}}{\tilde{\sigma}_{0}\cos\theta
}\left(  1-\exp\left(  -\frac{\tilde{\sigma}_{0}\cos\theta}{\tilde{V}%
_{0}\tilde{\tau}}\right)  \right)  -\frac{1}{2}\right]  \right)  \right\}
\nonumber\\
+\tilde{\sigma}_{0}(1-\cos\theta)=\Delta\tilde{\gamma}
\label{dimensionless final eq}%
\end{gather}

In the limiting cases of $\tilde{V}_{0}\gg1$ and $\tilde{\tau}\gg1$, Eq.
(\ref{final_2}) gives%
\begin{equation}
\frac{1}{2}\frac{\sigma_{0}^{2}}{E_{\infty}}\left(  1-\cos^{2}\theta\right)
+\sigma_{0}(1-\cos\theta)=\frac{\Delta\gamma}{d}
\label{high velocity dimensional}%
\end{equation}
which clearly differs from Eq. (\ref{final}), showing that the energy
dissipation due to frictional sliding at the interface is proportional to
$\frac{1}{2}\left(  \sigma_{0}^{2}/E_{\infty}\right)  \cos^{2}\theta$, which
leads to much tougher peeling behavior at small peeling angle, as the peeling
stress $\sigma_{0}$ diverges as $\sigma_{K}/\theta$. This result has been
already observed in Refs. \cite{Jagota2011,Begley2013} for purely elastic
tapes ($E$ is replaced by $E_{\infty}$), and it can be interpreted as the
emergence of an infinitely tough peeling behavior. Incidentally, it is worth
noticing that ultra-tough peeling has been also predicted to occur when the
tape is elastic and the substrate viscoelastic
\cite{Afferrante2016,Pierro2020}.

Similarly, in the limiting case of $\tilde{V}_{0}\tilde{\tau}\ll1$ with
$\tilde{V}_{0}\gg1$ (i.e. $V_{0}\gg d/t_{c}$), Eq. (\ref{final_2}) becomes%
\begin{equation}
\frac{1}{2}\frac{\sigma_{0}^{2}}{E_{0}}\left(  1-\cos^{2}\theta\right)
-\frac{\left[  \sigma_{0}\left(  1-\cos\theta\right)  \right]  ^{2}}{2E_{1}%
}+\sigma_{0}(1-\cos\theta)=\frac{\Delta\gamma}{d},
\label{plateau low velocity bis}%
\end{equation}
where the tape response in the adhered portion subjected to frictional shear
stresses is governed by the low frequency viscoelastic modulus $E_{0}$.
However, in Eq. (\ref{plateau low velocity bis}), the additional term
\begin{equation}
\frac{\sigma_{0}^{2}\left(  1-\cos\theta\right)  ^{2}}{2E_{1}}=\frac
{D_{\mathrm{f}}}{V_{0}bd}=\left(  1-\cos\theta\right)  ^{2}\frac
{D_{\mathrm{s}}}{V_{0}bd} \label{step dissipation}%
\end{equation}
takes into account for the viscoelastic energy dissipation per unit time
$D_{\mathrm{f}}$, triggered by the stress step change $\Delta\sigma=\sigma
_{0}-\sigma_{0}\cos\theta$, which still occurs at the peeling front. Indeed,
Eq. (\ref{viscoelastic energy dissipation 1}) is still valid provided that
$\sigma_{0}$ is replaced by $\Delta\sigma$. Notice that, as already discussed
before, for $\tilde{V}_{0}$ $\ll1$ (i.e. $V_{0}\ll d/t_{c}$) the term
$D_{\mathrm{f}}$ must also vanish, as even in the peeling section the tape
behaves elastically with modulus $E_{0}$. Therefore, for $\tilde{\tau}\ll1$
and $\tilde{V}_{0}$ $\ll1$ (i.e. $V_{0}\ll d/t_{c}$), Eq. (\ref{final_2})
becomes%
\begin{equation}
\frac{1}{2}\frac{\sigma_{0}^{2}}{E_{0}}\left(  1-\cos^{2}\theta\right)
+\sigma_{0}(1-\cos\theta)=\frac{\Delta\gamma}{d}%
\end{equation}
which holds true for purely elastic tapes (with elastic modulus $E=E_{0}$) in
the presence of interfacial frictional sliding (see Refs
\cite{Jagota2011,Begley2013}).\begin{figure}[ptbh]
\begin{center}
\centering\subfloat[\label{fig3a}]{\includegraphics[height=55mm] {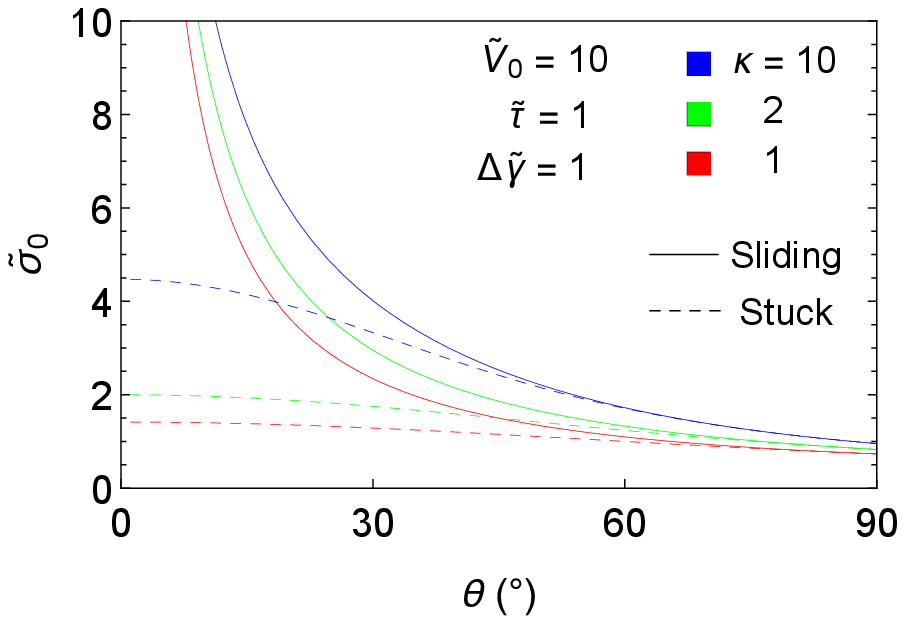}}
\hfill\subfloat[\label{fig3b}]{\includegraphics[height=55mm]{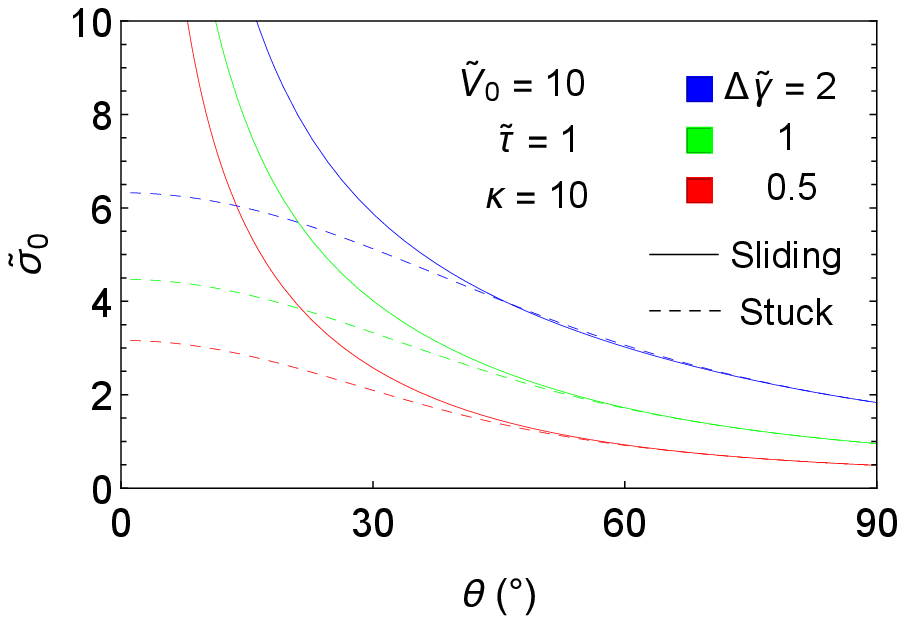}}
\end{center}
\caption{The dimensionless peeling stress $\tilde{\sigma}_{0}$ as a function
of the peeling angle $\theta$, for different values of (a) the viscoelasticity
parameter $\kappa=E_{\infty}/E_{0}$; and (b) the dimensionless energy of
adhesion $\Delta\tilde{\gamma}$. The dashed curves refer to the case of stuck
interface between the tape and the rigid substrate, whereas continuous curves
refer to frictionally sliding interfaces.}%
\label{fig3}%
\end{figure}

Figures \ref{fig3} show the dimensionless peeling stress $\tilde{\sigma}_{0}$
as a function of the peeling angle $\theta$, for both stuck and sliding
interfaces and different values of the parameter $\kappa=E_{\infty}/E_{0}$
[Fig. \ref{fig3a}], and of the energy of adhesion $\Delta\gamma$ [Fig.
\ref{fig3b}].

As already discussed, in the stuck case (dashed lines in both figures) we
recover the well-known elastic Kendall's solution, where the elastic modulus
is replaced by the high-frequency viscoelastic modulus $E_{\infty}$ [see Eq.
(\ref{final})]. Given the values of the low-frequency viscoelastic modulus
$E_{0}$, the peeling angle $\theta$ and work of adhesion $\Delta\gamma$, the
peeling force increases with the parameter $\kappa=E_{\infty}/E_{0}$.

On the other hand, in case of frictional sliding at the interface a very
different scenario emerges. This time, the peeling process is governed by Eq.
(\ref{final_2}), which, regardless of the $\kappa$ value, leads to unbounded
peeling forces for vanishing peeling angle $\theta$ (see continuous lines in
Figures \ref{fig3a} and \ref{fig3b}). In this case, the dimensionless peeling
stress obeys the equation $\tilde{\sigma}_{0}=\sqrt{2\Delta\tilde{\gamma}%
}/\theta$ for $\theta\rightarrow0$. Interestingly, such a result is in
agreement with several experimental observations on the peeling behavior of
insects pads in the presence of relative frictional sliding between the
fibrils and the substrate \cite{Labonte2016,Autumn2006b,Collino2014}. Figure
\ref{fig3b} presents the effect of the dimensionless energy of adhesion
$\Delta\tilde{\gamma}$ on the peeling behavior. As expected, regardless of the
specific interface behavior, increasing $\Delta\tilde{\gamma}$ leads to an
overall tougher peeling behavior, as the necessary stress $\tilde{\sigma}_{0}$
to sustain the peeling process increases \cite{Kendall1975}.

\begin{figure}[ptbh]
\begin{center}
\includegraphics[width=0.5\textwidth]{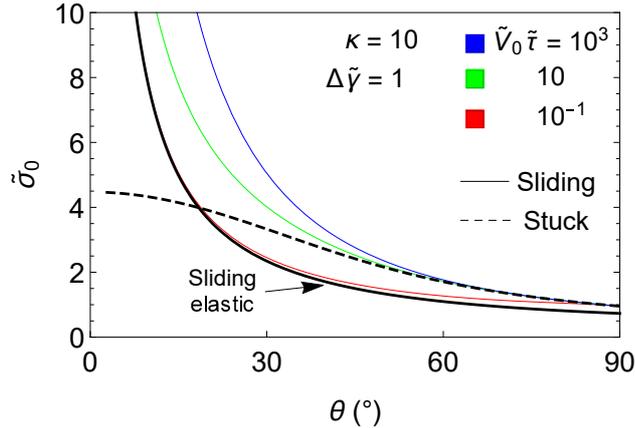}
\end{center}
\caption{The dimensionless peeling stress $\tilde{\sigma}_{0}$ as a function
of the peeling angle $\theta$, for different values of the dimensionless
parameter $\tilde{V}_{0}\tilde{\tau}$. The dashed curve refers to the case of
stuck interface between the tape and the rigid substrate, whereas continuous
curves refer to frictionally sliding interfaces. In the same figure, we also
plot for comparison the peeling behavior of an elastic tape in frictional
sliding. Reults refer to $\tilde{V}_{0}>>1$.}%
\label{fig4}%
\end{figure}

Figure \ref{fig4} shows the dimensionless peeling stress $\tilde{\sigma}_{0}$
as a function of the peeling angle $\theta$. This time, different values of
the dimensionless parameter $\tilde{V}_{0}\tilde{\tau}$ are considered. In the
same figure, we also report the purely elastic solution in the presence of
frictional sliding (with elastic modulus $E=E_{0}$). We observe that, for
relatively small values of the parameter $\tilde{V}_{0}\tilde{\tau}$ (red
curve) and moderately large peeling angles $\theta$, the value of
$\tilde{\sigma}_{0}$, observed in presence of frictional sliding, is lower
than the value predicted in the case of stuck interface (black dashed curve).
This is related to the different mechanisms of energy dissipation occurring in
each case. Indeed, for a stuck interface, the only source of energy
dissipation arises from the viscoelastic creep occurring in the detached
branch of tape (i.e. for $x>0$), which is independent on $\theta$. On the
contrary, when dealing with interfaces where frictional relative motion occurs
between the tape and the substrate, two additional sources of energy
dissipation can be identified: (i) the work done by the frictional shear
stress at the interface; and (ii) the viscoelastic creep occurring in the
portion of the tape adhered to the substrate and stretched by the interfacial
frictional shear stresses (i.e. for $-a\leq x<0$). However, for $\tilde{V}%
_{0}\tilde{\tau}\ll1$ and $0\ll\theta\lesssim\pi/2$, both these terms vanish
as the no viscoelastic creep occurs in the adhered sliding portion of the tape
(i.e. the tape response is governed by $E_{0}$, see Eq.
(\ref{plateau low velocity bis})) and the term $\cos^{2}\theta\rightarrow0$
(i.e. the work done by frictional shear stress is negligible, see Eq.
(\ref{friction})). Therefore, under these conditions, even in the case of
frictionally sliding interfaces, the only source of the energy dissipation is
the viscoelastic creep occurring in the detached tape, which can be quantified
as $D_{\mathrm{f}}$ through Eq. (\ref{step dissipation}). Since $D_{\mathrm{f}%
}=\left(  1-\cos\theta\right)  ^{2}D_{\mathrm{s}}<D_{\mathrm{s}}$, for
$\theta<\pi/2$, a lower peeling force is predicted in the frictional sliding
case compared to stuck interfaces, as indeed shown in Fig. \ref{fig4}.
Notably, the effect of the energy dissipation term $D_{\mathrm{f}}$ on the
overall peeling behavior can be appreciated by comparing the low speed
viscoelastic case (red curve) against the elastic limit (black continuous
curve). As already discussed in commenting Eq. (\ref{plateau low velocity bis}%
), a physical explanation of this phenomenon can be found by observing that,
in the case of frictional sliding interfaces, the step change occurring in the
tape stress at the peeling front is lower than in the case of stuck
interfaces, as in the former case the tape in the adhered portion close to the
peeling front is pre-stressed by the frictional shear stress by a quantity
$\sigma_{0}\cos\theta$. Thus, since for $\tilde{V}_{0}\tilde{\tau}<1$ the tape
pre-stress occurs at a very low excitation frequency (i.e. the tape response
does not present any creep), the resulting energy dissipation due to the
viscoelastic creep (only occurring in the detached strip) is smaller for
frictional sliding interfaces compared to the stuck case, in turn leading to
smaller peeling forces.\begin{figure}[ptbh]
\begin{center}
\includegraphics[width=0.5\textwidth]{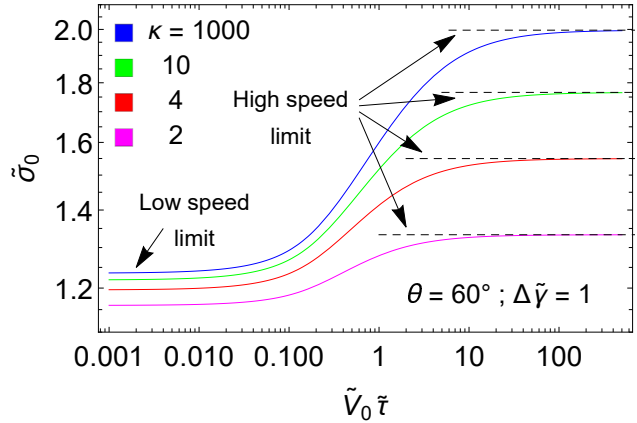}
\end{center}
\caption{The dimensionless peeling stress $\tilde{\sigma}_{0}$ as a function
of the dimensionless parameter $\tilde{V}_{0}\tilde{\tau}$, for different
values of the parameter $\kappa=E_{\infty}/E_{0}$. In the figure, both the
high and low speed plateau are highlighted. Reults refer to $\tilde{V}_{0}%
>>1$.}%
\label{fig5}%
\end{figure}

Figure \ref{fig5} shows the dimensionless peeling stress $\tilde{\sigma}_{0}$
as a function of the dimensionless parameter $\tilde{V}_{0}\tilde{\tau}$ for a
given value of $\theta$. All the cases refer to $\tilde{V}_{0}>1$. As expected
three different regimes can be observed depending on the value of $\tilde
{V}_{0}\tilde{\tau}$. For $\tilde{V}_{0}\tilde{\tau}\ll1$ an asymptotic
plateau for $\tilde{\sigma}_{0}$ is observed as predicted by Eq.
(\ref{plateau low velocity bis}), which depends on the value of $\kappa$. For
$\tilde{V}_{0}\tilde{\tau}\gg1$, the peeling behavior is governed by Eq.
(\ref{high velocity dimensional}) and depends on the high frequency
viscoelastic modulus $E_{\infty}$. Again, a plateau is observed for
$\tilde{\sigma}_{0}$, whose value saturates as for $\kappa\rightarrow\infty$
as the Rivlin's solution is approached in the case of infinitely stiff tapes.
At intermediate values of $\tilde{V}_{0}\tilde{\tau}$ the hysteretic
viscoelastic behavior of the tape plays a key role so that, in this transition
region, the peeling force increases with the peeling rate by following a power
law $\tilde{\sigma}_{0}\approx\left(  \tilde{V}_{0}\tilde{\tau}\right)  ^{n}$
where the exponent $n$ depends on the parameter $\kappa$.

\section{Conclusions}

In this study, we investigate the peeling behavior of a thin viscoelastic tape
peeled away from a rigid substrate. Specifically, we consider two alternative
scenarios: one, with the interface between the tape and the rigid substrate
under stuck adhesion (i.e. no sliding occurs); the other, assuming relative
sliding on a portion of the interface in the presence of frictional shear stresses.

We found that, in stuck interfaces, the overall viscoelastic peeling behavior
is independent of the peeling velocity, provided that the peeling velocity
$V_{0}\gg d/t_{c}$ (where $d$ is the thickness of the tape and $t_{c}$ is the
creep characteristic time of the viscoelastic material), and the peeling force
takes the value predicted by Kendall's peeling model with the elastic modulus
given by the high-frequency viscoelastic modulus $E_{\infty}$ of the tape
material. Under these conditions, the energy dissipation associated with the
viscoelastic creep of the tape is entirely\ localized in the detached portion
of the tape.

In the presence of frictional sliding at the interface additional sources of
energy dissipation come into play, which are associated with both the work
done by frictional shear stress and the viscoelastic hysteresis occurring in
the portion of the adhering tape subjected to frictional shear stresses. In
such conditions, the peeling force is predicted to continuously increase as
the peeling angle is decreased, leading to unbounded value for a vanishing
peeling angle. Also, the viscoelastic hysteretic behavior of the tape strongly
affects the dependence of the peeling force on the peeling velocity. Indeed,
for any given value of the peeling angle, three regions can be identified: (i)
the low velocity region, where a low plateau is reported for the peeling
force; (ii) the transition region, where the peeling force increases as a
power law of the peeling velocity, and (iii) the high velocity region, where a
high plateau of the peeling force occurs.

\begin{acknowledgement}
This project has received funding from the European Union's Horizon 2020
research and innovation programme under the Marie Sk\l odowska- Curie grant
agreement no. 845756 (N.M. Individual Fellowship). This work was partly
supported by the Italian Ministry of Education, University and Research under
the Programme \textquotedblleft Progetti di Rilevante Interesse Nazionale
(PRIN)\textquotedblright, Grant Protocol 2017948, Title: Foam Airless Spoked
Tire -- FASTire (G.C.)
\end{acknowledgement}

\end{document}